\documentstyle[12pt]{article}

\begin{document}
\def\thebibliography#1{\section*{REFERENCES\markboth
 {REFERENCES}{REFERENCES}}\list
 {[\arabic{enumi}]}{\settowidth\labelwidth{[#1]}\leftmargin\labelwidth
 \advance\leftmargin\labelsep
 \usecounter{enumi}}
 \def\newblock{\hskip .11em plus .33em minus -.07em}
 \sloppy
 \sfcode`\.=1000\relax}
\let\endthebibliography=\endlist

\hoffset = -1truecm
\voffset = -2truecm


\title{\large\bf
Markov-Yukawa Transversality On Covariant Null-Plane: Baryon Form Factor 
And  Magnetic Moments 
}
\author{
{\normalsize\bf
A.N.Mitra \thanks{e.mail: (1)ganmitra@nde.vsnl.net.in;
(2)anmitra@physics.du.ac.in}  and B.M. Sodermark*   
}\\
\normalsize 244 Tagore Park, Delhi-110009, India \\
\normalsize *High Energy Lab, Dept of Phys, Univ of Delhi,
Delhi-110007, India
} 

\maketitle

\begin{abstract}
The baryon-$qqq$ vertex function governed by the Markov-Yukawa
Transversality Principle ($MYTP$), is formulated via the Covariant 
Null-Plane Ansatz ($CNPA$) as a 3-body generalization of the corresponding 
$q{\bar q}$ problem, and employed to calculate the proton e.m. form factor 
and baryon octet magnetic moments.The e.m. coupling scheme is specified by 
letting the e.m. field interact by turn with the `spectator' while the 
two interacting quarks fold back into the baryon. The $S_3$ symmetry of 
the matrix element is preserved in all d.o.f.'s together. The $CNPA$ 
formulation ensures, as in the $q{\bar q}$ case, that the loop integral 
is free from the Lorentz mismatch disease of covariant instantaneity ($CIA$), 
while the simple trick of `Lorentz completion'ensures a Lorentz invariant
structure. The $k^{-4}$ scaling behaviour at large $k^2$ is reproduced.
And with the infrared structure of the gluonic propagator attuned to 
spectroscopy, the charge radius of the proton comes out at $0.96 fm$. 
The magnetic moments of the baryon octet, also in good accord with data, are
expressible as $(a+b\lambda)/(2+\lambda)$, where $a,b$ are purely 
geometrical numbers and $\lambda$ a  dynamics-dependent quantity.
\end{abstract}
                                                    
\newpage



\section*{1  Introduction}

There is a close connection between the physics of $q{\bar q}$ mesons
and that of $qqq$ baryons since in many ways a diquark behaves like a
color $3^*$ antiquark. Therefore  the dynamics of both systems should
be alike, so that it is fair to expect that a theory which is pre-tested 
for one should also work for the other, except for external characteristics 
like the number of degrees of freedom. Some important tests for any new 
form of dynamics are the scaling behaviour [1] of hadronic form factors, 
as well as the structure 
of baryon magnetic moments, the latter believed to be governed by chiral 
$SU(3)\times SU(3)$ symmetry [2] on the one hand, and vector meson 
dominance ($VMD$) [3] on the other. The form of dynamics we wish to 
consider here is the Markov-Yukawa Transversality Principle ($MYTP$) [4] 
via the covariant null plane ansatz ($CNPA$) [5], which is applicable to
all Salpeter-like equations [6]. It was recently applied to the e.m. form 
factor of the pion [5], showing the expected scaling behaviour ($\sim k^{-2}$).
With this check on the validity of $MYTP$ under $CNPA$ conditions [5],
we now wish to extend the same from the $q{\bar q}$ [5,6] to the $qqq$ dynamics.   
\par
	In Section 2 we derive the baryon-$qqq$ vertex function under 
$MYTP$ on the covariant null-plane ($CNPA$), on closely parallel lines to its Covariant Instantaneity ($CIA$) formulation [7,8], in the same notation [8] 
except when new features arise. Section 3 gives the matrix element for the 
baryon e.m. coupling wherein the spectator interacts with the external photon, 
and the two interacting quarks fold back into the baryon. Section 4 gives
the results for the e.m. form factor of the proton which confirm its $k^{-4}$ behaviour at large $k^2$, and gives a value of the charge radius at $0.96 fm$, 
with the infrared part of the gluon propagator [9] attuned to baryon spectroscopy [10]. Section 5 gives the corresponding results on the magnetic moments of the 
baryon octet within the same framework. Section 6 concludes with a summary.   

\section*{2  Derivation Of Baryon-$qqq$ Vertex Function}

\setcounter{equation}{0}
\renewcommand{\theequation}{2.\arabic{equation}}
   
To formulate the baryon-$qqq$ vertex function under $MYTP$, we proceed
as in ref.[8] for the 2- and 3-body kinematics, except for a generalization 
from covariant instantaneity ($CIA$) [7,8] to the covariant null-plane 
ansatz ($CNPA$) [5,6]. In this Section we  consider spinless quarks [8]
using the method of Green's functions, to be followed by the more realistic 
case of fermion quarks at the end of the Section. Now since the momentum kinematics for $q{\bar q}$ under $CNPA$ [5,6] turn out to be formally identical to those 
under the standard null-plane formalism [11,12] (which is easier to use), the latter notation ($\pm$) [11-12]  may be profitably employed instead of the $n_\mu$ dependent
notation [5,6]. Now the essential point of the $CNPA$ formalism [5,6] is that the
longitudinal $(z)$ and scalar $(0)$ components of a 4-momentum $p_{i\mu}$ for 
 quark $\#i$ in a hadron of mass $M$ and 4-momentum $P_\mu$ are [5,6]:
\begin{equation}\label{2.1}
p_{iz}; p_{i0} = \frac{M p_{i+}}{P_+}; \frac{M p_{i-}}{2 P_-}
\end{equation}
The former, along with the transverse components $p_{i\perp}$, obeys the 
`angular condition' [13,12] which effectively defines a 3-vector 
${\hat p}_i \equiv \{p_{i\perp}, p_{iz}\}$, a key ingredient for the $MYTP$ formulation to follow. 
   
\subsection*{2.1  3D-4D Interlinkage for $qq$ Green's Fns}

We first derive the 3D-4D interlinkage for the $qq$ system by the method 
of Green's functions, as a prototype for the $qqq$ system to follow. The 
Green's function under $MYTP$ satisfies the BSE [8]:
\begin{equation}\label{2.2}
(2\pi)^4 i G(q,q';P) = \frac{1}{\Delta_1 \Delta_2}\int d^4 q'' 
V({\hat q},{\hat q}'') G(q'',q';P)
\end{equation}
Now define the 3D Green's function [8]
\begin{equation}\label{2.3}
{\hat G}({\hat q},{\hat q}') =  \int {dq_0 dq_0'} G(q,q';P)
\end{equation} 
where the time-like components are defined a la (2.1). Integrating
both sides of (2.2) gives via (2.3) the 3D BSE for a $bound$ state
which does not need an inhomogeneous term:
\begin{equation}\label{2.4}
(2\pi)^3 D({\hat q}){\hat G}({\hat q},{\hat q}')  
=\int d^3 {\hat q}'' V({\hat q},{\hat q}'') {\hat G}({\hat q}'',{\hat q}')
\end{equation}
where the 3D denominator function $D({\hat q})$ is defined as 
\begin{equation}\label{2.5}
\frac{2i\pi}{D({\hat q})} = \int \frac{dq_0}{\Delta_1 \Delta_2}
\end{equation}
leading (for general unequal mass kinematics) to [12]
\begin{equation}\label{2.6}
D({\hat q}) = \frac{M}{P_+} D_+({\hat q}); \quad 
D_+({\hat q}) = 2P_+[{\hat q}^2- \frac{\lambda(M^2,m_1^2,m_2^2)}{4M^2}]
\end{equation}
where (2.1) defines the 3-vector ${\hat q}$ and $\lambda$ is the triangle 
function of its arguments.  Now define the hybrid Green's functions [8]:
\begin{equation}\label{2.7}
{\tilde G}({\hat q},q') = \int dq_0 G(q,q';P); \quad 
{\tilde G}(q,{\hat q'}) = \int dq_0' G(q,q';P)
\end{equation}      
Using (2.7) on the RHS of (2.2) gives
\begin{equation}\label{2.8}
(2\pi)^4 i G(q,q';P) = \frac{1}{\Delta_1 \Delta_2}\int d^3 {\hat q}'' 
V({\hat q},{\hat q}'') {\tilde G}({\hat q}'',q')
\end{equation}  
Integrating (2.2) w.r.t. $dq_0'$ only, and using (2.7) again, gives
\begin{equation}\label{2.9}
(2\pi)^4 i {\tilde G}(q,{\hat q}') = \frac{1}{\Delta_1 \Delta_2}\int 
d^3 {\hat q}'' V({\hat q},{\hat q}'') {\hat G}({\hat q}'',{\hat q}')
\end{equation}   
Eqs.(2.9) together with the 3D equation (2.4) for ${\hat G}$ gives
a connection between the hybrid ${\tilde G}$ and the 3D ${\hat G}$:
\begin{equation}\label{2.10}
{\tilde G}(q,{\hat q}') = \frac{D({\hat q})}{2i\pi \Delta_1 \Delta_2}
{\hat G}({\hat q},{\hat q}')
\end{equation}    
Interchanging $q$ and $q'$ in the last equation gives the dual result
\begin{equation}\label{2.11}
{\tilde G}({\hat q},q') = \frac{D({\hat q}')}{2i\pi \Delta_1' \Delta_2'}
{\hat G}({\hat q},{\hat q}')
\end{equation}       
Substituting these results in (2.8) gives the desired 3D-4D interconnection
\begin{equation}\label{2.12}
G(q,q';P) = \frac{D({\hat q})}{2i\pi \Delta_1 \Delta_2} 
{\hat G}({\hat q},{\hat q}';P) \frac{D({\hat q}')}{2i\pi \Delta_1' \Delta_2'}
\end{equation}
Now making  spectral representations for the 4D and 3D Green's functions
on both sides of eq.(2.12) in the standard manner [8], viz.,
\begin{equation}\label{2.13}
G(q,q';P) = \sum_n \Phi_n(q;P) \Phi^* (q';P)/ (P^2 +M^2);
\end{equation}
\begin{equation}\label{2.14}
{\hat G}({\hat q},{\hat q}') = \sum_n \phi_n({\hat q})\phi_n^*({\hat q}')
/(P^2+M^2)
\end{equation}
where $\Phi_n$ and $\phi_n$ are 4D and 3D wave functions respectively,
one can directly read off from (2.12) their interconnection, valid near 
a bound state pole (dropping the suffix $n$ for simplicity):
\begin{equation}\label{2.15}
\Gamma({\hat q}) \equiv \Delta_1 \Delta_2 \Phi(q;P)=
\frac{D({\hat q})\phi({\hat q})}{2i\pi}
\end{equation}
which tells us that the vertex function $\Gamma$ under $CNPA$ is again
a function of ${\hat q}$ only [8], except for its definition (2.1) under
$CNPA$. This derivation is a prototype for the $qqq$ case to follow.            
 
\subsection*{2.2 \quad 3D Reduction for Scalar $qqq$ BSE }

For the $qqq$ problem, we have a pair of internal variables which may be 
chosen in one of 3 distinct ways. With index $\#3$ as basis, the pair 
$\xi_3,\eta_3$ may be defined as [8]
\begin{equation}\label{2.16}
\sqrt{3} \xi_3 = p_1-p_2 ; \quad 3\eta_3 = -2p_3+p_1+p_2; \quad
P=p_1+p_2+p_3 
\end{equation}
The space-like and time-like parts of $\xi,\eta$ are defined as in (2.1), 
so that, e.g., 
$$ \xi_{z3} = \frac{M\xi_+}{P_+}; \quad \xi_{03}=\frac{M\xi_-}{2P_-} $$
and similarly for $\eta$. We shall also use the $\pm$ notation in parallel 
with the 3-vector notation in the following.   
\par
	The  Green's function, after taking out an overall $\delta$-
function for the c.m. motion, may be written as $G(\xi\eta;\xi'\eta')$ 
at the 4D level, while the fully 3D Green's function may be defined as [8]
\begin{equation}\label{2.17} 
{\hat G}({\hat \xi}{\hat \eta};{\hat \xi}'{\hat \eta}') =
\int {d\xi_0 d\eta_0 d\xi_0' d\eta_0'} G(\xi\eta;\xi'\eta')
\end{equation}
Both $G$ and ${\hat G}$ are $S_3$ symmatric, since the measure $d\xi_0 d\eta_0$ 
is $S_3$-invariant. In addition, two hybrid Green's functions are [8]:
\begin{equation}\label{2.18}   
{\tilde G}_{3\xi}({\hat \xi}_3 \eta_3;{\hat \xi}'_3 \eta'_3) =
\int {d\xi_{30} d\xi_{30}'} G(\xi\eta;\xi'\eta'); \quad
{\tilde G}_{3\eta}(\xi_3{\hat \eta}_3;\xi_3'{\hat \eta}_3') =
\int {d\eta_{30} d\eta_{30}'} G(\xi\eta;\xi'\eta'); 
\end{equation}
where the suffixes $3\xi$,$3\eta$ signify that ${\tilde G}$ 
is $not$ $S_3$ symmetric since the integration now involves only $one$ 
of the two $\xi$,$\eta$ variables.  Now the 4D $qqq$ BSE under $MYTP$  is [8]:  
\begin{equation}\label{2.19}
i(2\pi)^4 G(\xi\eta;\xi'\eta') = \sum_{123} \int \frac{9 d^4\xi''}
{16\Delta_1\Delta_2} V({\hat \xi}_3,{\hat \xi}_3'' 
G(\xi_3''\eta_3;\xi_3'\eta_3')
\end{equation}
where the factor $9/16$=$[\sqrt{3}/2]^4$ stems from the relation 
$2 q_{12} = \sqrt{3}\xi_3$, and the association  of $\eta_3$ with
$\xi_3''$ in the Green's function signifies that the spectator $\#3$
remains unaffected by the interaction in the $(12)$ pair, and so on
by turns cyclically. The 3D reduction is achieved by integrating (2.19)
via (2.17) which gives, as in the $CIA$ derivation [8]:  
\begin{equation}\label{2.20}
(2\pi)^3 {\hat G}({\hat \xi}{\hat \eta};{\hat \xi}'{\hat \eta}') 
=\sum_{123} \frac{3\sqrt{3}}{8 D_{12}} \int d^3{\hat \xi}_3''
V({\hat \xi}_3,{\hat \xi}_3'')    
{\hat G}({\hat \xi}_3''{\hat \eta}_3'';{\hat \xi}_3'{\hat \eta}_3')
\end{equation}
where, as in (2.6), $D_{12}$ = $\frac{M D_{12+}}{P_+}$, with 
\begin{equation}\label{2.21}
D_{12+} = 2\omega_{1\perp}^2 p_{2+}+ 2\omega_{2\perp}^2 p_{1+}-2 P_{12-}p_{1+}p_{2+}  
\end{equation}
and $\omega_{i\perp}^2 = m_i^2 + p_{i\perp}^2$; $P_{12} = p_1+p_2$.  Now making 
use of the on-shellness ($\Delta_3 =0$) of the spectator ($\#3$), we have
$ P_{12-}=P_- p_{3-}$ = $P_- \omega_{3\perp}^2 / p_{3+}$,   
whose substitution in (2.21) gives rise to the $S_3$ symmetric result:
\begin{equation}\label{2.22}
p_{3+} D_{12+} \equiv D_{++} = 2\sum_{123} p_{2+} p_{3+}\omega_{1\perp}^2
-2p_{1+} p_{2+} p_{3+} P_- 
\end{equation}
Substitution of this result in (2.20) gives rise to the requisite
3D BSE in which the denominator function may be identified with 
$D_{++}$ in an $exact$ fashion. This is a much neater result than
under $CIA$ [7] where a corresponding denominator function could
be obtained only through an approximate treatment [10]. Since this
quantity will play a crucial role in this study, we recast it in 
terms of the $\xi, \eta$ variables by first redifining it as  
$D_{++} \equiv P_+^2 D$, and $D \equiv D_0 + {\delta D}$, where  
\begin{equation}\label{2.23}
D_0 = \frac{1}{3}(\xi_\perp^2 + \eta_\perp^2) + \frac{1}{2}(1-\sum_{123}m_i^2/M^2)
(\xi_l^2 + \eta_l^2) + \frac{2}{3} (\sum_{123}m_i^2/3- M^2/9)
\end{equation}
with $(\xi_l, \eta_l) \equiv M(\xi_+,\eta_+)/P_+$ and
\begin{equation}\label{2.24}
{\delta D} = \frac{\eta_l}{2M}[\eta_l^2-3\xi_l^2] - \frac{3}{2M^2}
[\eta_l^2 \xi_\perp^2 + \xi_l^2 \eta_\perp^2]
\end{equation}
Note that this null plane description gives different scales for the 
longitudinal and transverse components of the $\xi, \eta$ variables.      
\par
	The $D$-function is the driving term for the 3D eq.(2.20)
which in turn  is the right vehicle for spectroscopy [10], and is the
source of the 3D wave function $\phi$ via the spectral representation
(2.14). In this respect, the main role is played by the $D_0$ 
function, while ${\delta D}$ is a correction term. 
   
\subsection*{2.3  Reconstruction of 4D Vertex Function}

We now indicate the steps for reconstruction of $G$ in terms of 3D ingredients.
 First, the hybrid function ${\tilde G}_{3\eta}$, eq.(2.18), is
 expressed in terms of fully 3D quantity ${\hat G}$ exactly as
in eq.(2.11) for the two-body problem:
\begin{equation}\label{2.25}
{\tilde G}_{3\eta}(\xi_3{\hat \eta}_3; \xi_3'{\hat \eta}_3')
=\frac{P_+ D}{2i\pi p_{3+}\Delta_1 \Delta_2}   \\
{\hat G}({\hat \xi}{\hat \eta};{\hat \xi}'{\hat \eta}')
 \frac{P_+ D'}{2i\pi p_{3+}'\Delta_1' \Delta_2'} 
\end{equation}
 In a similar way the fully 4D $G$ function is expressible in terms
of the hybrid function ${\tilde G}_{3\xi}$ as
\begin{equation}\label{2.26}
G(\xi\eta;\xi'\eta') = \sum_{123} \frac{P_+ D}{2i\pi p_{3+}\Delta_1 \Delta_2} 
{\tilde G}_{3\xi}({\hat \xi}_3\eta_3; {\hat \xi}_3' \eta_3')             
\frac{P_+ D'}{2i\pi p_{3+}'\Delta_1' \Delta_2'}
\end{equation}
At this stage we need an ansatz [8] on the ${\tilde G}_{3\xi}$ function
which, as in the $CIA$ case [8], is not determined from $MYTP$ $qqq$ dynamics:
\begin{equation}\label{2.27}
{\tilde G}_{3\xi}({\hat \xi}_3 \eta_3; {\hat \xi}_3'  \eta_3')
 = {\hat G}({\hat \xi}{\hat \eta}; {\hat \xi}' {\hat \eta}') F(p_3, p_3')
 \end{equation}
where we have incorporated the $S_3$ symmetry of ${\hat G}$ and taken the 
balance of the $p_3$ (spectator) dependence in the (unknown) $F$ function.
This is subject to an explicit self-consistency check for the ansatz (2.27)
which may be found by integrating both sides w.r.t. $dp_{3-} dp_{3-}'$,to give  
$$ \int \int \frac{M^2 dp_{3-} dp_{3-}'}{4 P_-^2} F(p_3,p_3') = 1 $$
This condition is satisfied by the ansatz [8]:
\begin{equation}\label{2.28}
F(p_3, p_3') = \frac{A_3}{\Delta_3} 
\delta[\frac{Mp_{3-}}{2P_-} - \frac{Mp_{3-}'}{2P_-}]   
\end{equation}
if $A_3$ is determined by the equation
$$ A_3 \int \frac{M dp_{3-}}{2P_- \Delta_3} = 1$$
which gives 
$$ A_3 = \frac{2M p_{3+}}{i\pi P_+}; \quad 
p_{3-} = \frac{\omega_{3\perp}^2}{p_{3+}} $$
After a little simplification, we have finally:
\begin{equation}\label{2.29}
F(p_3,p_3') =  4p_{3l}^2 \frac{\delta(\Delta_3)}{i\pi \Delta_3}
\end{equation}
which finally defines the 4D $G$ function in terms of ${\hat G}$ via the
sequence (2.27) and (2.26). Finally the spectral representations (2.13-14) near
a bound state pole give the connection between the $\#3$ part $\Phi_3$
of the 4D wave function and the 3D wave function $\phi$:
\begin{equation}\label{2.30}
\Phi_3 = \frac{2M D}{2i\pi \Delta_1 \Delta_2} \phi(\xi\eta) 
\sqrt{\frac{\delta(\Delta_3)}{i\pi \Delta_3}}
\end{equation}
whence the baryon-$qqq$ vertex function $V_3$ is inferred via
$$\Phi_3 \equiv \frac{V_3}{\Delta_1 \Delta_2 \Delta_3}: $$ 
\begin{equation}\label{2.31}
V_3 = \frac{M D\phi(\xi\eta)}{i\pi}{\sqrt{\frac{\Delta_3 \delta(\Delta_3)}
{i\pi}}}  
\end{equation}
\par
	For explanation on the appearance of the $\delta$-function under 
radicals in eq.(2.31), see ref.[8] where it has been shown that this has
nothing to do with any lack of connectedness in a 3-body amplitude.    

\subsection*{2.4   Baryon-$qqq$ Vertex with Fermion Quarks}

For the more realistic case of fermion quarks, we employ the method 
of Gordon reduction [14-15] whose logic and advantages have been 
described elsewhere in the context of the $q{\bar q}$ problem [5,6]. 
To extend the same to the 3-body case, define a (fictitious) scalar
function $\Phi$ which is related to the actual BS wave function 
$\Psi$ by [15]
\begin{equation}\label{2.32}
\Psi = \Pi_{123} S_{Fi}^{-1}(-p_i) \Phi (p_i p_2 p_3)
\end{equation}
with an explicit indexing w.r.t. the individual quarks, which however
can be subsumed in a common Dirac matrix space a la Blankenbecler 
et al [16], as illustrated in the next Section. The connection with 
sect.2.3 is now established by identifying $\Phi$ of (2.32) with the 
sum $\Phi_1+\Phi_2+\Phi_3$ where $\Phi_3$ is given by (2.30). Therefore
the form (2.31) for $V_3$ continues to be valid, except that its 
relation to $\Psi$ is
\begin{equation}\label{2.33}
\Psi = \Pi_{123} S_{Fi}(p_i) [V_1 + V_2 + V_3]
\end{equation}       
We end this Section with a listing of the (gaussian) structure of the 
3D wave function $\phi$ as a solution of the fermionic counterpart of
eqs.(2.20-24). Since this paper is not concerned with baryon 
spectroscopy (see [10] for details),we list merely the gaussian form of $\phi$:
\begin{equation}\label{2.34}
\phi = \exp{[-\frac{\xi_\perp^2 + \eta_\perp^2}{2\beta_t^2}
 -\frac{M^2 x^2 +M^2 y^2}{2\beta_l^2}]}
\end{equation}
where the transverse and longitudinal scale parameters follow from the
structure of the $D$ function (2.23), by proceeding as in [12,10]:  
\begin{eqnarray}\label{2.35}
\beta_t^4  &=& \frac{8M}{81}\omega_{qq}^2 /[1/4 -\frac{3C_0 \omega_{qq}^2}
{9M \omega_0^2}] \\  \nonumber
\beta_l^4  &=& \frac{8M}{81}\omega_{qq}^2 /[1/2 -\frac{3C_0 \omega_{qq}^2}
{9M \omega_0^2} -\frac{3 m_q^2}{2 M^2}]
\end{eqnarray}    
where the input parameters $\omega_{qq}$ etc are listed in [10,12]. The
numerical values of the $\beta^2$ parameters for the full baryon octet 
in $GeV^2$ units are
\begin{eqnarray}\label{2.36}
\beta_t^2(N)      &=& 0.068; \quad \beta_l^2 (N) = 0.054;  \\  \nonumber
\beta_t^2(\Sigma) &=& 0.080; \quad \beta_l^2(\Sigma)= 0.062; \\ \nonumber
\beta_t^2(\Lambda)&=& 0.076; \quad \beta_l^2(\Lambda)=0.061; \\  \nonumber
\beta_t^2(\Xi)    &=& 0.079; \quad \beta_l^2(\Xi) = 0.063.
\end{eqnarray}
  
\section*{3  E.M. Coupling Of $qqq$ Baryon}

The e.m. coupling of the $qqq$ baryon is given by fig.1 plus 
two more obtained by cyclic permutations of the indices. It shows that the spectator
($\#3$) scatters against the (space-like) photon before being 
re-absorbed into the baryon, while the two interacting quarks ($1,2$) fold back into the baryon.Fig.1 is in keeping with the standard additivity principle
(the hallmark of the quark model) for single quark transitions. On the other hand the (complementary) diquark-photon diagram [17], which does not show a similar
property, will be presumed to be $dynamically$ suppressed, hence left out of further consideration in this paper. 

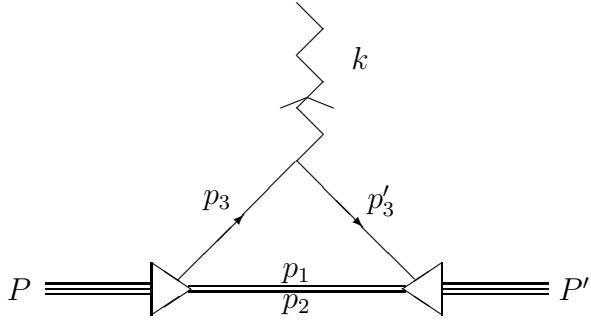
\begin{figure}[h]

\vspace{0.5in}

\begin{picture}(450,50)(-50,100)
\put (95,59.5){\line(1,0){40}}
\put (95,61.5){\line(1,0){40}}
\put (95,63.5){\line(1,0){40}}

\put (150,61.5){\line(-3,-2){14.5}}
\put (150,61.5){\line(-3,2){14.5}}
\put (135,71.5){\line(0,-1){20}}

\put (245,59.5){\line(1,0){40}}
\put (245,61.5){\line(1,0){40}}
\put (245,63.5){\line(1,0){40}}


\put(85,61.5){\makebox(0,0){$P$}}
\put(295,61.5){\makebox(0,0){$P'$}}
\put(160,95){\makebox(0,0){$p_3$}}



\put (230,61.5){\line(3,-2){14.5}}
\put (230,61.5){\line(3,2){14.5}}
\put (245,71.5){\line(0,-1){20}}

\multiput(190,110)(0,20){3}{\line(1,1){10}}
\multiput(200,120)(0,20){3}{\line(-1,1){10}}
\put (184,130){\line(5,2){10}} 
\put (194,134){\line(5,-2){10}}

\put(214,149){\makebox(0,0){$k$}}

\put(190,67.5){\makebox(0,0){$p_1$}}
\put(190,55.5){\makebox(0,0){$p_2$}}

\put (190,110){\vector(1,-1){25}} 
\put (214.5,85.2){\line(1,-1){20}}

\put (145,64.7){\vector(1,1){25}}
\put (169.5,89.5){\line(1,1){20}}

\put(222,95){\makebox(0,0){$p'_3$}}


\put (149,60.5){\line(1,0){82}}
\put (149,62.5){\line(1,0){82}}

\end{picture}
\vspace{1.0in}
\caption{Triangle loop for baryon e.m. vertex}

\end{figure}

\subsection*{3.1  Structure of E.M. Matrix Element}

\setcounter{equation}{0}
\renewcommand{\theequation}{3.\arabic{equation}}

The imbedding of the indices $123$ in a common Dirac matrix space
may be done a la ref.[16], after taking account of the $S_3$ symmetry
a la ref.[17]. In a $[2,1]$ representation of $S_3$ symmetry, the two 
spin functions $\chi', \chi''$ are expressible as [16,17]
\begin{eqnarray}\label{3.1}
\mid \chi'>  &=& [\gamma_5 \frac{C}{\sqrt{2}}]_{\alpha \beta} \otimes
U_\gamma (P)  \\  \nonumber
\mid \chi''> &=& [i{\hat \gamma}_\mu \frac{C}{\sqrt{6}}]_{\alpha \beta}
\otimes [\gamma_5 {\hat \gamma}_\mu U(P)]_\gamma
\end{eqnarray}
where ${\hat \gamma}_\mu$ would naively be expected to be transverse
to $P_\mu$, but one must again anticipate a problem analogous to the
Lorentz mismatch problem  associated with $CIA$ [7] which gave rise to 
unwarranted complexities [18] in form factors, and necessitated the 
alternative $CNPA$ [5] formulation for the $orbital$ matrix elements. 
In the context of $spin$ matrix elements this pathology shows up as high 
powers of $k^2$ in the concerned amplitudes if ${\hat \gamma}_\mu$ is 
defined as $ \gamma_\mu - \gamma.P P_\mu/P^2$ [7], which is equally unacceptable. 
On the other hand, the $CNPA$ [5] offers an alternative solution wherein 
the transversality is defined w.r.t. the (more universal) null-plane.       
The simplest possibility in this regard is to define
\begin{equation}\label{3.2}
{\bar \gamma}_\mu = \theta_{\mu\nu}\gamma_\nu; \quad
{\bar \gamma}_\mu^* = \theta_{\mu\nu}^* \gamma_\nu
\end{equation}
where
\begin{eqnarray}\label{3.3}
\theta_{\mu\nu} &=& \delta{\mu\nu} - n_\mu {\tilde n}_\nu = \theta_{\nu\mu}*; 
\\  \nonumber    
\theta_{\mu\mu} &=& 3; \quad \theta_{\mu\lambda}\theta_{\nu\lambda}^* 
= \theta_{\mu\nu} 
\end{eqnarray} 
These properties are consistent with those expected of a projection operator. 
Next, the $SU(6)$ operator for the baryon e.m. interaction is
\begin{equation}\label{3.4}
\Gamma_\mu = \sum_{1}^{3} \gamma_\mu^{(i)} \frac{e}{2} [\lambda_3^{(i)}
+ \frac{1}{\sqrt{3}} \lambda_8^{(i)}]
\end{equation}
The matrix elements of this operator must be taken between the
the $SU(6)$ wave functions [17] 
\begin{equation}\label{3.5}
W(P) = \frac{1}{\sqrt{2}}(\chi' \phi' + \chi'' \phi'')
\end{equation}
where the isospin functions $\phi', \phi''$ have matrix elements 
expressible in a common (baryon) basis as [17]
\begin{eqnarray}\label{3.6}
<\phi''\mid 1; {\bf \tau}^{(i)} \mid \phi''> &=& [1;-{\bf \tau}/3]\\ \nonumber 
<\phi'\mid 1; {\bf \tau}^{(i)} \mid \phi'>   &=& [1 ;{\bf \tau}]     
\end{eqnarray}     
Using (3.6), the matrix element of (3.4) between nucleon states $|N>$ is expressible as (c.f. [17]):
\begin{eqnarray}\label{3.7}
<N|\Gamma_\mu|N> &=& \int d\tau V_3^* V_3 N_B^2 [\frac{e}{2}<1+3\tau_z> 
[<\chi'\mid \gamma_\mu^{(3)} \mid \chi'> \\  \nonumber
                 & &  +\frac{e}{2}<1-\tau_z><\chi''\mid \gamma_\mu^{(3)} \mid \chi''>]
\end{eqnarray}
$N_B$ is a normalization factor to be defined further below.The two spin matrix elements are expressible in a factorized form as
\begin{eqnarray}\label{3.8}
<\chi'\mid \gamma_\mu^{(3)}\mid \chi'>  &=& T'\times A_\mu' 
\\   \nonumber    
<\chi''\mid \gamma_\mu^{(3)}\mid \chi''>&=& T_{\nu\nu'}''\times 
A_{\mu\nu\nu'}'';
\end{eqnarray}      
\begin{eqnarray}\label{3.9}
T'            &=& Tr[{\tilde S}_F(p_2)\frac{C^{-1}}{\sqrt{2}}\gamma_5 
S_F(p_1) \gamma_5 \frac{C}{\sqrt{2}}]  \\  \nonumber
T_{\nu\nu'}'' &=& Tr[{\tilde S}_F(p_2)\frac{C^{-1}}{\sqrt{6}}
{\bar \gamma}_{\nu'} S_F(p_1) {\bar \gamma}_\nu^*\frac{C}{\sqrt{6}}];
\end{eqnarray}
\begin{eqnarray}\label{3.10}
A_\mu'          &=&{\bar U}(P') S_F(p_3') i\gamma_\mu S_F(p_3) U(P) \\  \nonumber
A_{\mu\nu\nu'}''&=&{\bar U}(P') {\bar \gamma}_{\nu'}^*\gamma_5 S_F(p_3') 
i\gamma_\mu S_F(p_3) {\bar \gamma}_\nu \gamma_5 U(P) 
\end{eqnarray}
The symbol $\int d\tau$ in (3.8) stands for:
\begin{equation}\label{3.11}
\int d{\tau} \equiv \int d^4q_{12} d^4{\bar p}_3 
\end{equation}
where ${\bar p}_3$ =$(p_3+p_3')/2$.  

\subsection*{3.2  Baryon Normalization}

The evaluation of the spin matrix elements (see Appendix) reduces (3.7) to 
\begin{equation}\label{3.12}
<N|\Gamma_\mu|N> = \frac{e}{2} \int d\tau' W_3^* W_3' N_B^2 
[M_\mu'(1+3\tau_z)  + M_\mu'' (1-\tau_z)]
\end{equation}        
where $M_\mu', M_\mu''$ are given in eqs.(A.6, A.11) respectively,and 
$W_3$ is the reduced form of $V_3$, eq.(2.31) as under: 
\begin{equation}\label{3.13}
W_3 = \frac{M D\phi(\xi\eta)}{(\pi)^{3/2}}  
\end{equation}
Hera (see Appendix for details) the reduced measure $d\tau'$ is given by (A.7). 
Repeated use of Gordon reduction (A.8) leads finally to the `Sachs' form     
\begin{equation}\label{3.14}
<N|\Gamma_\mu|N> = e{\bar U}(P') [ F(k^2) \frac{{\bar P}_\mu}{M} + G(k^2)\frac{\sigma_{\mu\nu} k_\nu}{2M}] U(P)
\end{equation}
The baryon normalization now comes entirely from the first term in the limit of 
$k =0$. To fix this quantity,instead of demanding unit charge for the $proton$,
an asymmetric recipe heavily weighted against the neutron (see Eq.(3.12)), we
resort to a more symmetrical treatment between the two  by demanding the 
conservation of $baryon$ number, via $\omega$-like coupling [3],which is
equivalent to the conservation of $isoscalar$ charge.Thus the normalization 
condition boils down to 
\begin{equation}\label{3.15}
\frac{P_\mu}{M i (2\pi)^4} = \frac{N_B^2}{2} \int d\tau' [M_\mu'+M_\mu'']
\end{equation}
which is obtained from (3.12) after dropping the isovector parts. For later
purposes we define a parameter $\lambda$:
\begin{equation}\label{3.16}
\lambda = \frac{<{\bar \eta}^2 -3{\bar \xi}^2>}{M^2/3+m_3^2-{\delta m}^2}< 0 
\end{equation}
which arises from certain terms of $M_\mu''$, eq. (A.11), without a counterpart
from $M_\mu'$, eq.(A.6). This parameter will play a crucial role in the
determination of magnetic moments (see Section 5).     

\section*{4  Calculation Of The E.M. Form Factor}

\setcounter{equation}{0}
\renewcommand{\theequation}{4.\arabic{equation}}

The charge and magnetic form factors of the baryon are given by the functions
$F(k^2)$ and $G(k^2)$ in eq.(3.14) which in this model have $identical$ shapes.
Further,eq.(3.15) for normalization ensures that $F(0)\equiv 1$, while $G(0)$
gives directly the baryon magnetic moments in `baryon magnetons'($e/2M$). The
form factor problem is considered in this Section, followed by magnetic moments
in the next Section. To evaluate (3.12) in a $closed$ form, first write its
integration measure in detail as
\begin{equation}\label{4.1}
d^2 q_\perp dq_+ \frac{d{\bar q}_-}{2} d^4 {\bar p_3} \delta(\Delta_3).
\end{equation}
In the first step, integrate over $dq_-/2$ to give
\begin{equation}\label{4.2} 
\int \frac{d{\bar q}_-}{2} \frac{DD'}{\Delta_1 \Delta_2}     
= (2i\pi) \frac{p_{3+}}{4{\bar P}_+^2}(D+D') 
\end{equation}
The next step is to integrate over the factors $dp_{3-}\delta(\Delta_3)/2$ 
in (4.1) to give $\frac{1}{2 p_{3+}}$. Combining (4.1-2), the net measure becomes:
\begin{equation}\label{4.3} 
d\tau_1 = (2i\pi) d^2 \xi_\perp d^2\eta_\perp \frac{M^2 dx dy (D+D')}{8};
\quad x;y = \frac{\xi_+;\eta_+}{{\bar P}_+}
\end{equation}
Next, define a common basis ${\bar \xi}$ and ${\bar \eta}$ as follows:
\begin{equation}\label{4.4}
\eta, \eta' = {\bar \eta} \mp k/3; \quad  \xi = \xi' = {\bar \xi},
\end{equation} 
in terms of which the product of the 3D wave functions becomes
\begin{equation}\label{4.5}
\phi \phi' = \exp {[-\frac{\xi_\perp^2+\eta_\perp^2}{\beta_t^2} - 
\frac{f(x,y)}{\beta_l^2}]}; 
\end{equation} 
\begin{eqnarray}\label{4.6}
2f(x,y)       &=& \sum_{\pm}[\frac{M^2 x^2}{(1 \pm {\hat k}/2)^2}
+\frac{M^2 (y \mp {\hat k}/3)^2}{(1 \pm {\hat k}/2)^2}] \\  \nonumber 
x,y;{\hat k}  &=& \frac{({\bar \xi}_+, {\bar \eta}_+;k_+)}{{\bar P}_+}
\end{eqnarray}
Giving the translation 
\begin{equation}\label{4.7}
x \rightarrow x;  \quad y \rightarrow y - 2\sigma_k;\quad 
\sigma_k \equiv \frac{{\hat k}^2/6}{1+{\hat k}^2/4}
\end{equation} 
the function $f(x,y)$ reduces to
\begin{equation}\label{4.8}
f(x,y) = \frac{(M^2 x^2 + M^2 y^2)(1+{\hat k}^2/4)}{(1-{\hat k}^2/4)^2}
  + 2M^2 \sigma_k/3
\end{equation}    
The same translation to the ${\bar D}=(D+D')/2$ function, dropping odd terms, gives 
\begin{eqnarray}\label{4.9}
{\bar D}(k^2)  &=& \frac{\xi_\perp^2 + \eta_\perp^2}{3} + 
\frac{2}{3}m_q^2 -\frac{2}{27} M^2 \\  \nonumber
               & & + \frac{(M^2-3m_q^2)}{2}[\frac{(x^2+y^2)(1+{\hat k}^2/4)}
{(1-{\hat k}^2/4)^2} + 2\sigma_k/3] + ``R''; \\  \nonumber 
 ``R''         &=&  -\frac{3}{2} M^2 y^2 \frac{\sigma_k}{(1-{\hat k}^2/4)^2} 
 +{\cal O}({\hat k}^4) 
\end{eqnarray}   
The rest of the integration is now routine gaussian for casting the e.m. 
matrix element in the form (3.14). Using the basic formulae
\begin{equation}\label{4.10} 
\int d^2\xi_\perp d^2\eta_\perp \exp{[-\frac{\xi_\perp^2+\eta_\perp^2}
{\beta_t^2}]} = (\pi \beta_t^2)^2; 
\end{equation}
\begin{equation}\label{4.11}
 \int Mdx Mdy \exp{[-f(x,y)/\beta_l^2]} = \pi \beta_l^2
\frac{(1-{\hat k}^2/4)^2}{1+{\hat k}^2/4} 
\exp{[-\frac{2M^2\sigma_k}{3\beta_l^2}]}
\end{equation}        
and other allied results, all integrations are carried out explicitly,
and the form factors $F,G$ identified a la (3.14). The common form is
\begin{equation}\label{4.12}
F(k^2) = \frac{(1-{\hat k}^2/4)^2}{1+{\hat k}^2/4} 
\exp{[-\frac{2M^2\sigma_k}{3\beta_l^2}]} \frac{{\bar D}(k^2)}{{\bar D}(0)}         
\end{equation}

\subsection*{4.1  Results on Proton Form Factor}

Eq.(4.12) represents our final formula for the proton form factor.
Before comparison with experiment [19]
however, we need to invoke the principle of `Lorentz completion' [5] to give
$F(k^2)$ an explicitly Lorentz-invariant look. The trick is to consider
a collinear frame so that $P_\perp = P_\perp' =0$. From this frame it is
easy to see that [5]
\begin{equation}\label{4.13} 
{\hat k}^2 = \frac{4 k^2}{4 M^2 + k^2};\quad 1-{\hat k}^2/4 =
\frac{4M^2}{4M^2+k^2} 
\end{equation} 
Substitution of (4.13) in (4.12) gives an explicitly  Lorentz invariant result.
Eq.(4.12) then shows that $F(k^2) \sim k^{-4}$ for large $k^2$, in
conformity with the scaling law [1] for the baryon.    
\par
	Next we consider the e.m. radius of the proton which is given
by the formula
\begin{equation}\label{4.14}
<R^2> = -6 \partial_{k^2} F(k^2)|_{k^2 =0}
\end{equation}        
Expanding (4.12) up to ${\cal O}(k^2)$ and collecting the coefficients
of the indicated derivative gives
\begin{equation}\label{4.15}
<R^2> = -\frac{6}{M^2}(-3.39) = 23.14 GeV^{-2} = (0.96 fm)^2
\end{equation} 
on substitution of the $\beta^2$ values from (2.36). These 
results are in fair accord with the observed value of $\sim 0.90 fm$
for the proton's e.m. radius [20], considering the fact that the
parameters are not adjustable but attuned to $qqq$ spectroscopy [10]. 

\section*{5  Baryon Magnetic Moments}

\setcounter{equation}{0}
\renewcommand{\theequation}{5.\arabic{equation}}

Before calculating the magnetic moments (as the coefficients of $\sigma_\mu$), we first generalize the formula (3.12) for the full baryon octet, with the replacements $(1+3\tau_z)\rightarrow f'$ and $(1-\tau_z) \rightarrow f''$, where the latter are listed in Table 1 below. Although we are now allowed to set $k^2=0$, we must take 
account of the i) unequal mass kinematics; and ii) the normalization (3.15) for 
the general baryon case.  Unequal mass kinematics is ensured 
simply by the replacement $ 3 \rightarrow \sum_{123} $ with an 
appropriate change of $S_3$ basis by keeping track of the index 
$\#3$ in the terms $m_3$ and ${\delta m}^2$ appearing in the 
quantities $M_\mu'$ and $\mu''$.                

\begin{center}
\Large \bf
Table I: Flavour Factors $f'$ and $f''$ for Baryons \\
\large
\begin{tabular}{|r|l|c|c|}
\hline
Baryon type&$f'$&$f''$ \\
\hline
$N$&$\frac{e(1+3\tau_z)}{4}$&$\frac{e(\tau_z-1)(1+\lambda)}{12}$ \\
$\Sigma$&$\frac{e(1+3T_z)}{4}$&$\frac{e(1-T_z)(1+\lambda)}{12}$ \\
$\Lambda$&$\frac{-e}{4}$&$\frac{-e(1+\lambda)}{12}$ \\
$\Xi$&$\frac{-e}{2}$&${\bar \tau}_z\frac{-e(1+\lambda)}{6}$ \\
$\Lambda-\Sigma$&$\frac{e\sqrt{3}}{4}$&$\frac{e\sqrt{3}(1+\lambda)}{12}$ \\
\hline
\end{tabular}
\end{center}

The flavour factors in Table 1 are mostly geometrical [17], except for the 
(small) parameter $\lambda$  which enters the expression for $f''$ 
in the second column of the table. As noted at the end of Section 3, the origin 
of this term may be traced to the last two terms of $M_\mu''$, eq.(A.11), which 
may be regarded as dynamical corrections to $SU(6)$ that affect $M_\mu''$ but
not $M_\mu'$. The parameter $\lambda$ which represents this effect, is already
expressed by eq.(3.16), where the $<...>$ sign stands for the effect of 
integration over the internal variables a la Section 4. The ratio $\lambda$ also enters the normalization via the RHS of eq.(3.15), which contributes a factor
$(2+\lambda)^{-1}$ to the magnetic moment. Collecting all these results, 
the baryon magnetic moments in baryon magneton($B$) units are all expressible as
\begin{equation}\label{5.1}
\mu_{Bm} = \frac{a + b\lambda}{2 +\lambda}
\end{equation}
where $a,b$ are geometrical numbers given in Table 2 below, along with the
results (in baryon magnetons) of this calculation, experiment [19] and the
Schwinger model [2].  
\begin{center}
\Large \bf
Table 2: $a,b$ values and magnetic moments  \\
\large
\begin{tabular}{|r|l|c|c|c|c|}
\hline
Baryon &$a$&$b$&$\mu_B$&Expt[19]&Sch [2] \\
\hline
$p$&$4$&$0$&$+2.710$&$+2.793$&$+2.42$ \\
$n$&$-8/3$&$-2/3$&$-1.570$&$-1.913$&$-1.62$ \\
$\Lambda$&$-4/3$&$-1/3$&$-0.741$&$-0.614$&$-0.614$ \\
$\Lambda-\Sigma$&$4/\sqrt{3}$&$1/\sqrt{3}$&$+1.278$&$1.61$&...\\
$\Sigma^+$&$4$&$0$&$+2.407$&$+2.33$&$+2.36$ \\
$\Sigma^-$&$-8/3$&$-2/3$&$-1.456$&$-0.89$&$-0.87$ \\
$\Xi^0$&$-8/3$&$-2/3$&$-1.438$&$-1.236$&$-1.356$ \\
$\Xi^-$&$-4/3$&$+2/3$&$-0.876$&$+0.75$&$-0.55$ \\
\hline
\end{tabular}
\end{center}

The results for the magnetic moments are in fair accord with experiment[19]  
as well as with the Schwinger model of e.m. substitution [2] in accordance
with $VMD$ [3]. The following symmetry relations, $if$ the results are expressed in baryon magneton units, may also be noted:
\begin{equation}\label{5.2}
\mu_p =\mu_{\Sigma^+}; \quad \mu_n=\mu_{\Sigma^-}= \mu_{\Xi^0}=
2\mu_\Lambda = \frac{-2}{\sqrt 3}\mu_{\Lambda-\Sigma^0}
\end{equation}  

\section*{6. Summary And Conclusion}

In this paper, we have attempted to extend the 
Markov-Yukawa Transversality Principle [4] on the covariant null-plane 
from the $q{\bar q}$ problem [5] to the closely related $qqq$ system,  
and given an explicit construction of the corresponding baryon-$qqq$ 
vertex function.As a test of this Principle, its applications have been
carried out on two allied quantities viz., i) the e.m. form factor of the 
proton; and ii) the magnetic moments of the baryon octet. The calculation 
of the former has been carried out on the lines of the corresponding work 
on the meson e.m. form factor [5], using the method of `Lorentz completion' for obtaining an explicitly Lorentz-invariant result. Not only is the scaling 
law [1] reproduced, but also a value of the proton e.m. radius obtained in 
fair accord with the data [20], when the infra-red part of the gluon propagator 
is attuned to hadron spectroscopy [10]. The application to the baryon magnetic 
moments reveals an interesting set of symmetry relations, eq.(5.2), when 
expressed in units of `baryon magnetons'. The results also show  a good pattern 
of accord with experiment[19] as well as with the Schwinger model [2]. These
results may be regarded more by way of `calibration' of a relatively  new 
principle,(the $MYTP$ [4]), than any attempt at exploring newer aspects of 
these familiar quantities, such as the gluonic radius of the proton [21], 
higher order e.m. effects on the proton radius [22], strangeness effects [23], 
etc, for which the interested reader is referred to other publications [24]. 
 
\par
	One of us (BMS) is grateful to Prof.R.K.Shivpuri for the
hospitality of the High Energy Lab of Delhi Univ.  

\section*{Appendix A: Evaluation Of Spin Matrix Elements}

\setcounter{equation}{0}
\renewcommand{\theequation}{A.\arabic{equation}}
  
We indicate here the main steps for evaluating the spin matrix elements in 
the factorized forms (3.8-10), leading to the form (3.12). Consider first
the $T'$ and $A_\mu'$ terms. Taking the traces, $T'$ simplifies to 
$$ T' = 2\frac{m_1 m_2-p_1.p_2}{\Delta_1 \Delta_2}=
\frac{\Delta_1 + \Delta_2 - P_{12}^2-{\delta m}^2}{\Delta_1\Delta_2} \Rightarrow
\frac{-P_{12}^2-{\delta m}^2}{\Delta_1 \Delta_2} $$
where $P_{12}=p_1+p_2$, and we have used the result that in the 
null-plane formalism, the poles in the $\Delta{1,2}$ propagators are 
on opposite sides of the $q_-$ plane [12], so that the `virtualities'
$\Delta_{1,2}$ in the numerator effectively drop out. A further 
simplification arises from the result
$$ P_{12}^2= ({\bar P}-{\bar p}_3)^2 \Rightarrow  M^2- 2{\hat m}_3 M^2/3 + m_3^2 $$
where ${\hat m}_3\approx 1/3$ is the fraction of momentum carried by the spectator,
and some odd powers in $\eta_3$ have been dropped. Thus
\begin{equation}\label{A.1}
T' = \frac{M^2/3 + m_3^2}{\Delta_1 \Delta_2}
\end{equation}
The multiplying factor $A_\mu'$, eq.(3.10), may be written as
$$ A_\mu' = {\bar U}'(P') \frac{(m_3-i\gamma.p_3')i\gamma_\mu
(m_3-i\gamma.p_3)}{\Delta_3 \Delta_3'} U(P) $$
For further processing, we collect together the singular factors (S.F.)
involving $\Delta_3, \Delta_3'$ in $A_\mu'$ as well as the vertex 
functions $V_3, V_3'$, eq.{2.31), in the form
\begin{equation}\label{A.2}
S.F. \equiv \frac{\sqrt{\Delta_3 \delta(\Delta_3)}
\sqrt{\Delta_3' \delta(\Delta_3')}}{\Delta_3 \Delta_3'}
\end{equation}    
Since this singular function is non-vanishing only at a couple of 
points it can be bounded by making use of the inequalities
$$ h.m. < g.m. < a.m.$$
in a $compensatory$ manner :
$$ \sqrt{\Delta_3 \Delta_3'} \geq \frac{2 \Delta_3 \Delta_3'}
{\Delta_3+\Delta_3'}; $$
$$\sqrt{\delta(\Delta_3) \delta(\Delta_3')} \leq 
[\delta(\Delta_3) + \delta(\Delta_3')]/2 $$
Multiplying these two factors together and substituting in (A.2)
we finally obtain the result
\begin{equation}\label{A.3}
S.F. \approx \frac{\delta(\Delta_3) + \delta(\Delta_3')}
{\Delta_3 + \Delta_3'} \Rightarrow 
\frac{\delta(\Delta_3) - \delta(\Delta_3')}{\Delta_3 - \Delta_3'} 
\end{equation}
where the last step has made repeated use of the vanishing property 
of the argument of a $\delta$-function. The last expression in turn
is expressible as a $derivative$:
\begin{equation}\label{A.4}
\frac{\delta(\Delta_3) - \delta(\Delta_3')}{\Delta_3 - \Delta_3'}
\approx - \partial_{{\bar \Delta}_3} \delta({\bar \Delta}_3)
= -\partial_{m_3^2} \delta({\bar \Delta}_3)
\end{equation}
where ${\bar \Delta}_3$ $\approx$ $m_3^2 + {\bar p}_3^2 $.  
The trick is now to transfer the burden of differentiation from
the $\delta$-function to the rest of the integrand in the sense of
integration by parts. Then the differentiation w.r.t. $m_3^2$ boils
down to the expression
$$  +\partial_{m_3^2} {\bar U}(P') (m_3-i\gamma.p_3')
i\gamma_\mu(m_3-i\gamma.p_3) U(P)   
\Rightarrow  {\bar U}(P') [i\gamma_\mu(1 + \frac{M}{3m_3}) U(P) $$
on making use of the Dirac equation, and dropping a small $\eta)_3$ 
term. Collecting all these results we have
\begin{equation}\label{A.5}
A_\mu' \rightarrow {\bar U}(P') i\gamma_\mu(1 + \frac{M}{3m_3}) U(P)
\delta({\bar \Delta}_3)
\end{equation}
This quantity is multiplied by $T'$, eq.(A.1), and the product
of (A.1) and (A.5) represents one of the dual spin-matrix elements:
\begin{equation}\label{A.6}
M_\mu' = (1+\frac{M}{3m_3})\frac{M^2/3+m_3^2}{\Delta_1 \Delta_2}
{\bar U}(P')[\frac{{\bar P}_\mu}{M} + \sigma_\mu] U(P) 
\end{equation} 
where the $\delta$-function on the RHS of (A.5) is absorbed in the
integration measure (3.11) which is now redefined as
\begin{equation}\label{A.7}
\int d\tau' \equiv \int d^4q_{12} d^4{\bar p}_3 \delta(\Delta_3)
\end{equation}
and the Dirac matrix  $\gamma_\mu$ has been be Gordon-reduced as [14]
\begin{equation}\label{A.8}
i\gamma_\mu \rightarrow \frac{{\bar P}_\mu}{M} + \sigma_\mu; \quad  
\sigma_\mu \equiv  \frac{i\sigma_{\mu\nu}k_\nu}{2M}     
\end{equation}
This Gordon reduction at the quark level determines the relative 
strengths of the charge and magnetic form factors in eq.(3.14) of text, 
in the Sachs convention. 
\par 
	In a similar way the pair $T''$ and $A_\mu''$ can be simplified, 
except for a bit heavier algebra stemming from the extra tensor 
indices involved in each, as well as the presence of ${\bar \gamma}_\nu$ 
and its covariant conjugate ${\bar \gamma}_\nu^*$ which are defined via 
eqs.(3.2-3). Using the properties (3.3) of the projection operators 
$\theta_{\mu\nu}$, it is not difficult to show that          
\begin{equation}\label{A.9}
T_{\nu\nu'}'' = \frac{\theta_{\nu\lambda}^* \theta_{\nu'\lambda'}}
{3\Delta_1 \Delta_2} [(M^2/3 + m_3^2) \delta_{\lambda\lambda'}    
+ {\bar \eta}_\lambda {\bar \eta}_{\lambda'} 
- 3{\bar \xi}_\lambda {\bar \xi}_{\lambda'}]
\end{equation}  
where the `bar' symbols are as defined in (3.2). The bar symbols also 
include the effect of averaging over the $\xi,\eta$ values for the
initial and final baryons. Similarly, the
quantity $A_{\mu\nu\nu'}''$ is treated exactly as in (A.5) to give
\begin{equation}\label{A.10}
A_{\mu\nu\nu'}'' = \theta_{\nu\rho}\theta_{\nu'\rho'}^* (1 + \frac{M}{3m_3}) 
  {\bar U}(P')\gamma_{\rho'} i\gamma_\mu \gamma_{\rho} U(P) 
\delta({\bar \Delta}_3)
\end{equation}
Thus the product of (A.9) and (A.10) defines the dual quantity to
$M_\mu'$ of (A.6), which on contracting over some tensor indices gives
\begin{equation}\label{A.11}
M_\mu'' = \frac{(1+\frac{M}{3m_3})}{\Delta_1 \Delta_2}
{\bar U}(P') [(\frac{{\bar P}_\mu}{M} - \frac{\sigma_\mu}{3})
(M^2/3+m_3^2) - ({\bar \eta}^2/3 - {\bar \xi}^2) \sigma_\mu] U(P) 
\end{equation}    
where again the $\delta$-function in $A_{\mu\nu\nu'}''$ of (A.10)
hhas been absorbed in the new integration measure (A.7).
The final formula for the spin matrix element in terms of 
$M_\mu', M_\mu''$ is given in eq.(3.12) of text.

\end{document}